\documentclass[12pt]{iopart}
\usepackage{verbatim}
\usepackage{color}
\usepackage{graphicx}

\begin{document}
\topical[]{Introduction to topological superconductivity and Majorana fermions}

\author{Martin Leijnse and Karsten Flensberg}

\address{Center for Quantum Devices, Niels Bohr Institute, University of Copenhagen, 
Universitetsparken 5, 2100~Copenhagen \O, Denmark }
\ead{leijnse@fys.ku.dk}
\begin{abstract}
This short review article provides a pedagogical introduction to the rapidly growing research field of Majorana fermions in 
topological superconductors.  
We first discuss in some detail the simplest "toy model" in which Majoranas appear, namely a one-dimensional 
tight-binding representation of a $p$-wave superconductor, introduced more than ten years ago by Kitaev. 
We then give a general introduction to the remarkable properties of Majorana fermions in condensed matter systems,
such as their intrinsically non-local nature and exotic exchange statistics, and explain why 
these quasiparticles are suspected to be especially well suited for low-decoherence quantum information processing. 
We also discuss the experimentally promising (and perhaps already successfully realized) possibility of creating 
topological superconductors using semiconductors with strong spin-orbit coupling,
proximity-coupled to standard $s$-wave superconductors and exposed to a magnetic field. 
The goal is to provide an introduction to the subject for experimentalists or theorists who are new to the field, focusing 
on the aspects which are most important for understanding the basic physics.
The text should be accessible for readers with a basic understanding of quantum mechanics and second quantization, 
and does not require knowledge of quantum field theory or topological states of matter.
\end{abstract}

\section{Introduction}
Topological superconductivity is an interesting state of matter, partly because it is associated with quasiparticle 
excitations which are Majorana fermions (MFs).
In particle physics, MFs are (fermionic) particles which are their own anti-particles~\cite{Wilczek09}. 
It is still unclear if there are elementary particles which are MFs, but they are likely to exist as quasiparticle 
excitations in certain condensed matter systems, where a MF is a quasiparticle which is its "own hole". 
The condensed matter version of MFs have attracted massive theoretical interest, mainly because of their 
special exchange statistics: They are non-abelian anyons~\cite{Stern10rev}, meaning that particle exchanges are non-trivial 
operations which in general do not commute. This is unlike other known particle types where an exchange operation merely 
has the effect of multiplying the wavefunction with +1 (for bosons) or -1 (for fermions) or a general phase factor
$\phi$ (for "ordinary" (abelian) anyons). Furthermore, a MF is in a sense half of a normal fermion, meaning that 
a fermionc state is obtained as a superposition of two MFs. 

It should be noted, however, that any fermion can be written as a combination of two MFs, which basically corresponds to splitting 
the fermion into a real and an imaginary part, each of which is a MF. Normally this is a purely mathematical operation without physical 
consequences, since the two MFs are spatially localized close to each other, overlap significantly, and cannot be addressed individually.
When we talk about MFs here, we mean that a fermionic state can be written as a superposition of two MFs which are \emph{spatially separated}
(or prevented from overlapping in some other manner). Such a highly delocalized fermionic state is protected from most 
types of decoherence, since it cannot be changed by local perturbations affecting only one of its Majorana constituents. 
The state can, however, be manipulated by physical exchange of MFs because of their non-abelian 
statistics, which has lead to the idea of 
low-decoherence \emph{topological} quantum computation~\cite{Nayak08rev}.

Being its own hole means that a MF must be an equal superposition of an electron and a hole state. It is natural to search for 
such excitations in superconducting systems, where the wavefunctions of Boguliubov quasiparticles have both an electron and a hole 
component. The most common type of superconducting pairing is of so-called $s$-wave symmetry, where Cooper pairs are formed 
of electrons with opposite spin projections (forming a singlet). 
In second quantization language, the annihilation operator of a Boguliubov quasiparticle in an $s$-wave superconductor has the form 
$b = u c^\dagger_\uparrow + v c_\downarrow$, 
where $c_\sigma$ annihilates a normal fermion with spin projection $\sigma = \uparrow,\downarrow$ (we neglect for simplicity to write 
out additional quantum numbers which are irrelevant here). 
Having \emph{equal} electron and hole components mean that MFs are instead associated with an annihilation operator 
of the type $\gamma = u c^\dagger_\sigma + u^* c_\sigma$, which 
is hermitian and therefore equal to the creation operator, $\gamma = \gamma^\dagger$.
Note that, in contrast to the $s$-wave Boguliubov quasiparticle operator, the fermion operators making 
up the MF have equal spin projections.
Such quasiparticles do not occur in most types of superconductors and  
were instead first predicted to occur in the $\nu = 5/2$ fractional quantum Hall 
state~\cite{Moore91}. However, as we will discuss in more detail below, isolated MFs occur in general in 
vortices and on edges of effectively spinless superconducting systems with 
triplet pairing symmetry~\cite{Kopnin91, Read00, Sengupta01, Ivanov01, Zhang08, Kraus09, Wimmer10, Sato09, Sato10} 
($p$-wave pairing symmetry in one dimension (1D) and $p_x \pm ip_y$ pairing symmetry in two dimensions (2D)). 
Triplet pairing has been predicted for the ground state of the superconductor 
Sr$_2$RuO$_4$~\cite{DasSarma06}, but is very sensitive to disorder and has never been observed experimentally. 
However, the existence of (spatially separated) MFs is a topological invariant~\cite{Hasan10} (hence the name 
\emph{topological} superconductors). As a result, they will exist in all 
systems with the same topological properties as a $p$-wave or $p_x \pm i p_y$-wave superconductor
(we will come back to the precise meaning of this later). 

A few years ago, the search for MFs took a big step forward when Fu and Kane~\cite{Fu08} showed that $p_x \pm ip _y$-wave-like 
pairing may also occur for the surface states of a strong topological insulator when brought into tunneling 
contact with an ordinary $s$-wave superconductor (giving rise to proximity-induced superconductivity~\cite{deGennes64, Doh05, vanDam06} 
in the topological insulator). 
The necessary underlying physical ingredient is the strong spin-orbit coupling of the topological insulator, giving rise to 
split bands with momentum-dependent spin directions. In addition, by coupling the topological insulator also to a magnetic 
insulator, the resulting induced Zeeman splitting lifts the Kramer's degeneracy and allows an effectively spinless regime to be reached.
However, as was realized shortly after the pioneering work of Kane and Fu, the strong spin-orbit coupling in 
certain two-dimensional semiconductor quantum wells should also do the job~\cite{Sau10, Alicea10, Akhmerov11}. 
Also here superconductivity can be induced through the proximity effect and magnetism either induced by a magnetic 
insulator~\cite{Sau10} or provided by an external magnetic field~\cite{Alicea10}.
Shortly thereafter, two works~\cite{Oreg10, Lutchyn10} suggested a further simplification by instead using 1D 
semiconducting wires.
There have also been proposals to create MFs e.g., in vortices in doped topological insulators~\cite{Hosur11}, 
on the interface between a ferromagnet and a superconductor deposited on a 
two-dimensional topological insulator~\cite{Nilsson08, Fu09, Linder10}, in cold atomic gases~\cite{Gurarie05, Tewari07}, 
in carbon nanotubes~\cite{Sau11a, Klinovaja12, Egger12}, and using chains of quantum dots~\cite{Sau11b}, just to name a few.

During the last couple of years, a number of experimental groups have taken up the challenge to create MFs. 
Very recently this quest may have seen success~\cite{Mourik12}, and several other groups have made observations which can be 
interpreted as signatures of Majoranas~\cite{Williams12, Rokhinson12, Deng12, Das12}. 
There is no doubt that these early findings will be scrutinized in future experiments and time will tell which 
are genuine observations of Majorana physics. 
In any case, once MFs have been conclusively produced and detected in the lab, the truly exciting experimental work begins, to 
test their theoretically predicted properties and to design setups for ever more advanced manipulation of the quantum information
they can encode.

In this short review article we give an introduction to the topic of topological superconductivity and MFs, which is aimed at both 
experimental and theoretical physicists without prior knowledge of the field. We explain the generic necessary 
ingredients of topological superconductivity, as well as the basic properties of MFs, and how they can be realized in standard semiconductors
proximity-coupled to $s$-wave superconductors. Two excellent review articles discussing the subject have appeared 
recently~\cite{Alicea12rev, Beenakker11rev}. Therefore, we do not attempt to give an exhaustive review of the subject, or to provide detailed derivations 
of all results, but instead focus on providing physical insight into what we feel are the most important basic concepts. 
We also try to convey the excitement over MFs by briefly discussing some of their exotic properties, such as the non-abelian 
statistics and the resulting potential for topological quantum computation. 
The text should be accessible for all readers with a basic knowledge of quantum mechanics and the second quantization formalism.

\section{Majorana fermions in $p$-wave superconductors}
We start our discussion by introducing a simple Hamiltonian, describing a spinless $p$-wave superconductor, which has eigenstates 
which are spatially isolated MFs. 
It is most intuitive to start from a 1D tight-binding chain with $p$-wave superconducting pairing,
as first introduced by Kitaev~\cite{Kitaev01}, described by the Hamiltonian 
\begin{eqnarray}\label{eq:kitaevH}
	\mathcal{H}_\mathrm{chain} = -\mu \sum_{i=1}^N n_i - \sum_{i=1}^{N-1} \left( t c_i^\dagger c_{i+1} + \Delta c_i c_{i+1} + h.c. \right),
\end{eqnarray}
where $h.c.$ means hermitian conjugate, $\mu$ is the chemical potential, $c_i$ is the electron annihilation operator for site $i$, 
and $n_i = c_i^\dagger c_i$ is the associated number operator. The superconducting gap, $\Delta$, and hopping, $t$, 
are assumed to be the same for all sites. 
We can then choose the superconducting phase $\phi$ to be zero, such that $\Delta = |\Delta|$.
Note that time-reversal symmetry is broken in Eq.~(\ref{eq:kitaevH}) since we only consider one value for the spin projection, 
i.e., effectively spinless electrons (we suppress the spin label). Furthermore, the superconducting pairing is non-standard since it 
couples electrons with the same spin (in contrast to standard $s$-wave pairing, which only couples electrons with 
opposite spin projection). Note also that electrons on neighboring sites are paired: The sites cannot be doubly 
occupied by the spinless electrons because of the Pauli exclusion principle.

\begin{figure}[h!]
  \includegraphics[height=0.32\linewidth]{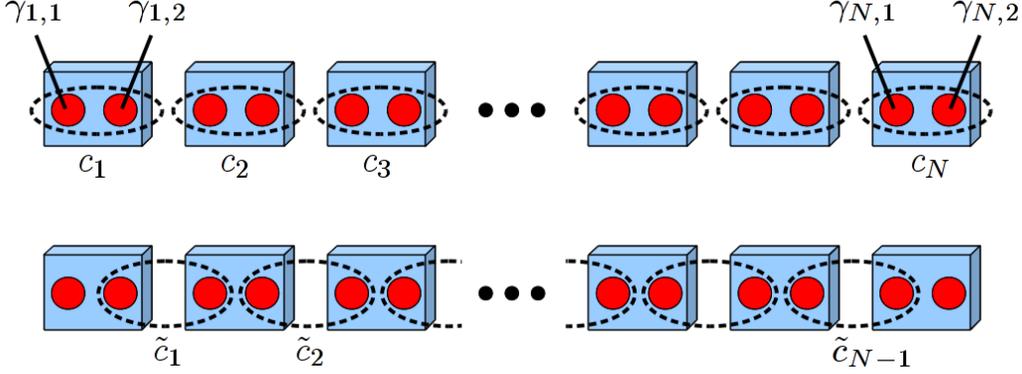}
\begin{centering}	
	\caption{\label{fig:kitaev}
	Sketch of Kitaev's 1D $p$-wave superconducting tight binding chain. 
	Upper panel: The fermion operators on each site $i$ of the chain can be split into two Majorana operators, 
		   $\gamma_{i,1}$ and $\gamma_{i,2}$.
	Lower panel: In the limit $\mu = 0$, $t = \Delta$, the Hamiltian is diagonal in fermion operators which are 
		   obtained by combing instead Majorana operators on neighboring sites, $\gamma_{i+1,1}$ and $\gamma_{i,2}$.
		   This leaves two unpaired Majorana operators, $\gamma_{1,2}$ and $\gamma_{N,1}$, which can be combined to form 
		   one zero energy, highly non-local fermion operator, $\tilde{c}_M$.}
\end{centering}
\end{figure}

We now want to rewrite Eq.~(\ref{eq:kitaevH}) in terms of Majorana operators (we will see shortly why this is useful). 
It was mentioned above that MFs are basically obtained by splitting a fermion into its real and imaginary parts. 
Therefore we write
\begin{eqnarray}
\label{eq:kitaevMFannihilation}
	c_i = \frac{1}{2} \left( \gamma_{i,1} + i \gamma_{i,2} \right), \\
\label{eq:kitaevMFcreation}
	c_i^\dagger = \frac{1}{2} \left( \gamma_{i,1} - i \gamma_{i,2} \right),
\end{eqnarray}
where $\gamma_{i,j}$ are Majorana operators living on site $i$.
That they are indeed Majorana operators is seen by inverting 
Eqs.~(\ref{eq:kitaevMFannihilation})--(\ref{eq:kitaevMFcreation}), giving
\begin{eqnarray}
\label{eq:kitaevMFinv1}
	\gamma_{i,1} = c_i^\dagger + c_i, \\
\label{eq:kitaevMFinv2}
	\gamma_{i,2} = i \left( c_i^\dagger - c_i \right),
\end{eqnarray}
which are clearly hermitian and therefore Majorana operators.
Figure~\ref{fig:kitaev} shows a sketch of Kitaev's chain and the upper panel indicates how the fermion operators on each 
site are split into Majorana operators.

The Majorana physics is most easily understood when $\mu =0$, $t = \Delta$, in which case inserting 
Eqs.~(\ref{eq:kitaevMFannihilation})--(\ref{eq:kitaevMFcreation}) into the Hamiltonian~(\ref{eq:kitaevH}) results in
\begin{eqnarray}\label{eq:kitaevHMF}
	\mathcal{H}_\mathrm{chain} = -i t \sum_{i=1}^{N-1} \gamma_{i,2} \gamma_{i+1,1}.
\end{eqnarray}
In fact, Eq.~(\ref{eq:kitaevHMF}) is nothing but an alternative way of writing the diagonalized Hamiltonian. 
To see this we go back to a fermionic representation by noting that, analogous to 
Eqs.~(\ref{eq:kitaevMFannihilation})--(\ref{eq:kitaevMFcreation}), where a fermion 
on site $i$ was split into two Majorana operators living on site $i$, we can construct new fermion operators, $\tilde{c}_i$,
by combining Majorana operators on neighboring sites
\begin{eqnarray}
	\label{eq:kitaevnewfermion}
	\tilde{c}_i = ( \gamma_{i+1,1} + i \gamma_{i,2} ) / 2.
\end{eqnarray}
This pairing is demonstrated in the lower panel of Fig.~\ref{fig:kitaev}.
In terms of these new fermions we find $-i \gamma_{i,2} \gamma_{i+1,1} = 2 \tilde{c}_i^\dagger \tilde{c}_i = 2 \tilde{n_i}$
and therefore
\begin{eqnarray}\label{eq:kitaevHdiag}
	\mathcal{H}_\mathrm{chain} = 2 t \sum_{i=1}^{N-1} \tilde{c}_i^\dagger \tilde{c}_i.
\end{eqnarray}
Thus, $\tilde{c}_i$ are the annihilation operators corresponding to the eigenstates and the energy cost of 
creating a $\tilde{c}_{i}$ fermion is $2t$. 
The Majorana operators are merely a formal way of writing the Hamiltonian and the physical 
excitations are fermionic states at finite energy, obtained by a superposition of nearest neighbor MFs.

So far there appears to be nothing special about the diagonal Hamiltonian in Eqs.~(\ref{eq:kitaevHMF}) and~(\ref{eq:kitaevHdiag}).
However, the Majorana operators $\gamma_{N,2}$ and $\gamma_{1,1}$, which are localized at the two ends of the wire, 
are completely missing from Eq.~(\ref{eq:kitaevHMF})! 
These two Majorana operators can equivalently be described by a single fermionic state with operator 
\begin{eqnarray}\label{eq:kitaevfermion}
	\tilde{c}_M = ( \gamma_{N,2} + i \gamma_{1,1} ) / 2.
\end{eqnarray}
This is a highly non-local state since $\gamma_{N,2}$ and $\gamma_{1,1}$ are localized on opposite ends of the chain. 
Furthermore, since this fermion operator is absent from the Hamiltonian, occupying the corresponding state requires zero energy.
Thus, in contrast to "normal" superconductors, where the ground state is non-degenerate and consists of a superposition 
of even-particle-number states (condensate of Cooper pairs), the Hamiltonian~(\ref{eq:kitaevH}) allows for an odd number 
of quasiparticles at zero energy cost. The ground state is therefore two-fold degenerate, corresponding to having in total 
an even or odd number of electrons in the superconductor. This even or oddness, also called parity, corresponds to the 
eigenvalue of the number operator of the zero-energy fermion, $n_M = \tilde{c}_M^\dagger \tilde{c}_M = 0 (1)$  
for even (odd) parity. 

The above argument was made for the very special case $\Delta = t$ and $\mu = 0$, but one can show that the Majorana end 
states remain as long as the chemical potential lies within the gap~\cite{Kitaev01}, $|\mu| < 2t$. 
In the general case, however, the MFs are not completely localized only at the two edge sites of the wire, but decay exponentially away 
from the edges. The MFs remain at zero energy only if the wire is long enough that they do not overlap.

The Hamiltonians for the continuum version of a $p$-wave superconductor in 1D and 2D are  
\begin{eqnarray}\label{eq:pgap1d}
	\mathcal{H}^\mathrm{pw}_{\mathrm{1D}} &=& \int d x \; \left[ \Psi^\dagger (x) \left( \frac{p_x^2}{2 m} - \mu \right) \Psi (x) + 
				    \Psi (x) |\Delta| e^{i\phi} p_x \Psi(x) + h.c. \right], \\
	\label{eq:pgap2d}
	\mathcal{H}^\mathrm{pw}_{\mathrm{2D}} &=& \int d^2r \; \left[ \Psi^\dagger (\mathbf{r}) \left( \frac{\mathbf{p}^2}{2 m} - \mu \right) 
				     \Psi (\mathbf{r}) + \Psi (\mathbf{r}) |\Delta| e^{i\phi} 
				     \left( p_x \pm ip_y \right) \Psi(\mathbf{r}) + h.c. \right], \nonumber \\
\end{eqnarray}
where $\Psi^\dagger (\mathbf{r})$ is the real-space creation operator, $\mathbf{p}$ is the momentum, $m$ is the effective electron mass, 
and we have re-introduced the superconducting phase $\phi$.
As we saw above, in the 1D case MFs appear at the edges of the wire. They will also appear at transition points between topological and 
non-topological regions.
For example, if the chemical potential or the hopping amplitude varies along the wire and $|\mu| > 2t$ in some segment, two additional 
MFs will appear at the transition points where the gap closes~\cite{Oreg10}. 
 
Similarly, in a 2D $p_x \pm ip_y$-wave superconductor, MFs appear in vortices in the 
superconducting pairing potential~\cite{Read00, Ivanov01, Kraus09}. 
Alternatively, MFs can appear if the gap is closed by variations in the chemical potential or electrostatic potential~\cite{Wimmer10}.
It is worth noting that isolated MFs can appear even in a 
spinful $p_x \pm ip_y$-wave superconductor, in so-called half-quantum vortices~\cite{Ivanov01}, where there is a vortex for only one direction of 
the triplet (note that this also means a breaking of time-reversal symmetry). 

If there are in total an odd number of gap closings, an additional MF will appear somewhere in the system 
to guarantee that there is always an even number (exactly where this additional MF appears depends on the details, 
see Ref.~\cite{Alicea12rev} for a detailed discussion).

\section{Properties of Majorana fermions}
Before explaining how MFs can be realized in more realistic systems, we now discuss in more detail some of their 
generic properties, which are independent of the specific system in which they appear. 

\subsection{Representation in terms of fermionic operators}
Let us assume that we have a system with $2N$ spatially well-separated MFs, $\gamma_1, \ldots \gamma_{2N}$.
The number of MFs is necessarily even since one MF contains half the degrees of freedom of a normal fermion. 
Similar to the case of Kitaev's chain, the Majorana operators are obtained by 
splitting a normal fermion $f_i$ in its real and imaginary parts (cf., Eq.~(\ref{eq:kitaevfermion}))
\begin{eqnarray}\label{eq:fermion}
	f_i &=& (\gamma_{2i-1} + i \gamma_{2i})/2.
\end{eqnarray}
The inverse relation is then 
\begin{eqnarray}
\label{eq:inversefermion1}
	\gamma_{2i-1} &=& f^\dagger + f, \\
\label{eq:inversefermion2}
	\gamma_{2i} &=& i(f^\dagger - f).
\end{eqnarray} 
Obviously, the Majorana operators are hermitian, $\gamma_j = \gamma_j^\dagger$. Using the fermionic anti-commutation relations for the 
$f_i$-fermions, it is easily verified that the Majorana operators satisfy the anti-commutation relation
\begin{eqnarray} \label{eq:anticommutation}
	\{\gamma_i, \gamma_j\} = 2 \delta_{ij},
\end{eqnarray}
which is somewhat reminiscent of normal fermions. There are, however, important differences. 
From Eq.~(\ref{eq:anticommutation}), we see that $\gamma_i^2 = 1$. Thus, acting twice with a Majorana operator, we get back the 
same state we started with. There is therefore no Pauli principle for MFs (cf., $c^2 = (c^\dagger)^2 = 0$ for normal 
fermion operators $c$). 
In fact, we cannot even speak of the occupancy of a Majorana mode. We can try to count the occupancy by constructing a  
"Majorana number operator", $n_i^\mathrm{MF} = \gamma_i^\dagger \gamma_i$. 
However, using hermiticity, $\gamma_i^\dagger = \gamma_i$, together with $\gamma_i^2 = 1$, we find $n_{i}^\mathrm{MF} \equiv 1$.
Similarly, $\gamma_i \gamma_i^\dagger \equiv 1$. Thus, the Majorana mode is in a sense always empty and always filled and counting 
does not make any sense.

It would still be very useful to use some sort of number states. These are instead provided through the $f_i$ fermions as
we already saw above in the example with Kitaev's chain. 
Since these are normal fermions, there are number states $|n_1, \ldots, n_N\rangle$ which are eigenstates 
of the fermionic number operators, $n_i = f_i^\dagger f_i$, with eigenvalue $n_i = 0, 1$ (by Pauli exclusion principle). 
Note that the labelling of the $\gamma$'s, and thereby the pairing into fermionic states, is arbitrary and merely represents a 
choice of basis for the number states. However, if two MFs come close enough to overlap, it is natural to choose to combine them into 
a fermion. To describe an overlap, $t$, between $\gamma_{2i-1}$ and $\gamma_{2i}$, the only term one can introduce into the Hamiltonian is
\begin{eqnarray}\label{eq:overlap}
	\frac{i}{2} t \gamma_{2i-1} \gamma_{2i} &=& t \left(n_i - \frac{1}{2}\right),
\end{eqnarray}
which corresponds to a finite energy cost for occupying the corresponding fermionic state ($t>0$). 
If the MFs do not overlap, the groundstate is $2^N$-fold degenerate, corresponding to each $n_i$ being equal to zero or one.
The sum of all occupation numbers, $\sum_{i=1}^{N} n_i$, being even or odd now reflects whether the total number of electrons 
in the superconductor is even or odd (even or odd parity). To change the parity, electrons have to be physically added 
to or removed from the superconductor.

The experimentally measurable quantities are the occupation numbers $n_i$. Such a measurement
can be done by bringing two MFs close together and measuring the energy of the corresponding state~\cite{Kitaev03}, which 
reveals the occupation through Eq.~(\ref{eq:overlap}). One could also use interferometry~\cite{Bishara09} or coupling to 
conventional superconducting qubits~\cite{Hassler10}.
Note that it does not make sense to talk about the "state of a MF" since a single MF contains only "half a degree of freedom". 
The only physical observables are the fermionic occupation numbers.

\subsection{Non-abelian statistics}
It is a crucial ingredient for non-abelian statistics to have a degenerate groundstate, which is separated from all excited states 
by a gap (ideally the induced superconducting gap, but it could also be a smaller gap to for example finite-energy vortex- or edge 
states). Then adiabatic operations, such as the slow exchange of quasiparticle positions, can in principle bring the system from one
groundstate to another. It is, of course, not obvious that such a transformation indeed takes place, which will depend on the details 
of the system. In the case of MFs in a $p_x \pm ip_y$ superconductor, Ivanov~\cite{Ivanov01} provided a simple and elegant proof of 
the non-abelian statistics which we sketch here. 
(The Supplementary Information of Ref.~\cite{Alicea10b} provides a proof in the 
case of 1D wires, where MFs can be moved using closely spaced electronic "keyboard" gates and particle exchange is made possible 
by connecting the 1D wires in T-junctions.)

\begin{figure}[h!]
  \includegraphics[height=0.22\linewidth]{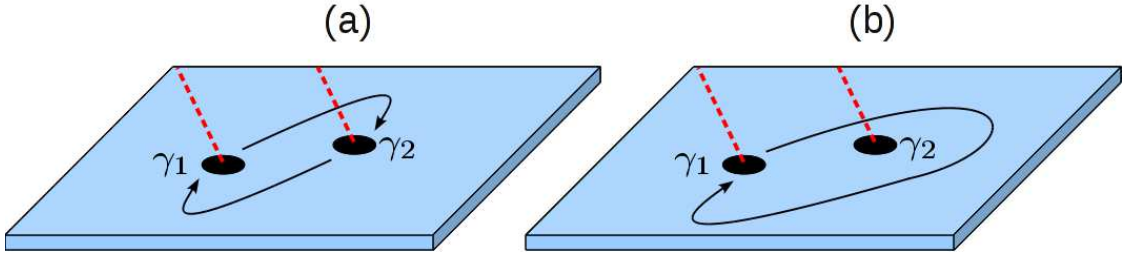}
\begin{centering}	
	\caption{\label{fig:braid}
	Sketch of vortices in a 2D $p_x \pm ip_y$ superconductor, hosting MFs described by the operators $\gamma_1$ and $\gamma_2$.
	Inside the vortex core the superconducting gap vanishes, $\Delta \rightarrow 0$, and going around the vortex the phase of 
	the superconducting condensate (Cooper pairs) increases by $2\pi$. 
	Therefore we introduce branch cuts emanating from the vortex cores (red dashed lines), where the phase makes a 
	$2\pi$ jump.
	(a) In a clockwise exchange, vortex one necessarily crosses the branch cut of vortex two.
	(b) When bringing vortex one around vortex two, both vortices cross the branch cut of the other vortex.}
\end{centering}
\end{figure}

Imagine that we have two vortices 
in a two-dimensional topological superconductor, hosting MFs described by the operators $\gamma_1$ and $\gamma_2$ at the vortex cores, 
see Fig.~\ref{fig:braid}. Each vortex is associated with a winding of $2\pi$ of the superconducting phase $\phi$. 
We can choose $\phi$ to be single-valued everywhere, except for at branch cuts (red dashed lines in Fig.~\ref{fig:braid}) 
emanating from each vortex, such that $\phi$ changes by $2 \pi$ when crossing this 
line (the direction of the branch cuts can be chosen arbitrarily). 
Vortices could perhaps be moved using the tip of a scanning tunneling microscope, or by local magnetic gates.
If we now exchange vortices one and two in a clockwise manner 
as indicated in Fig.~\ref{fig:braid}(a), vortex 1 crosses a branch cut and acquires a $2 \pi$ phase shift, while vortex 2 does not acquire a phase.
The superconducting phase is the phase of the Cooper pairs in the condensate. The MF in vortex 1, which is made up 
from single (rather than products of two) fermion operators, then acquires a phase of $\pi$ upon crossing the branch cut. 
The result of this exchange operation is thus
\begin{eqnarray}
\label{eq:braid1}
	\gamma_1 &\rightarrow& - \gamma_2,\\ 
\label{eq:braid2}
	\gamma_2 &\rightarrow& + \gamma_1.
\end{eqnarray}
This transformation is described by $\gamma_i \rightarrow B_{12} \gamma_i B_{12}^\dagger$, where the so-called braid operator
is given by
\begin{eqnarray}\label{eq:exchange}
	B_{12} &=& \frac{1}{\sqrt{2}} \left( 1 + \gamma_1 \gamma_2 \right).
\end{eqnarray}
This choice of operator is made unique by requiring that, in a system with more than two vortices (and therefore MFs), 
all the others are unaffected by the exchange of vortices one and two. Note that an anti-clockwise exchange instead results in 
$\gamma_1 \rightarrow  \gamma_2$, $\gamma_2 \rightarrow - \gamma_1$, which is described by the operator 
$\tilde{B}_{12} = ( 1 - \gamma_1 \gamma_2 ) / \sqrt{2}$. Of course, if we had chosen the branch cuts in a different direction the situation 
could be reversed, but this does not matter as long as we make a choice and stick with that.

Next, we want to discuss the effect of bringing vortex one around vortex two and back to its original position. Topologically, 
this is equivalent to two 
successive exchanges. Thus, the associated operator is given by $B_{12}^2 = \gamma_1 \gamma_2$, leading to the transformation
\begin{eqnarray}
\label{eq:exchangesquare1}
	\gamma_1 \rightarrow \left( \gamma_1 \gamma_2 \right) \gamma_1 \left( \gamma_1 \gamma_2 \right)^\dagger = -\gamma_1, \\
\label{eq:exchangesquare2}
	\gamma_2 \rightarrow \left( \gamma_1 \gamma_2 \right) \gamma_2 \left( \gamma_1 \gamma_2 \right)^\dagger = -\gamma_2. 
\end{eqnarray}
Bringing vortex one around vortex two thus has the effect of introducing a minus sign into each Majorana operator.
An alternative way of deriving Eqs.~(\ref{eq:exchangesquare1})--(\ref{eq:exchangesquare2}) is sketched in Fig.~\ref{fig:braid}(b). 
When bringing vortex one around vortex two, it necessarily crosses the branch cut of vortex two, but in addition forces 
vortex two to cross the branch cut of vortex one. Therefore, each Majorana operator acquires a phase shift of $\pi$.

We now go back to the case of vortex exchange. The effect of the braid operator acting on the number states is
\begin{eqnarray}
\label{eq:exchange0}
	B_{12} |0\rangle =  \frac{1}{\sqrt{2}} \left(1 + i \right) |0 \rangle, \\
\label{eq:exchange1}
	B_{12} |1\rangle =  \frac{1}{\sqrt{2}} \left(1 - i \right) |1 \rangle, 
\end{eqnarray}
where $|1\rangle = f_1^\dagger |0\rangle$, with $f_1 = (\gamma_1 + i \gamma_2)/2$ as discussed above. 
Thus, exchanging the two MFs has the rather mundane effect of multiplying the number states with a phase factor.

In fact, it is natural that the exchange operation cannot change the eigenvalue of the number operator in a system with 
only two MFs, since this encodes whether
there are in total an even or odd number of particles in the superconductor, a quantity which is not changed by particle exchanges.
To find non-trivial effects of exchange operations we must consider a system with at least four MFs, described in terms of the 
fermionic number states $|n_1 n_2\rangle$. 
Let us now investigate the effect of exchanging neighboring MFs, described by braid operators $B_{i, i+1}$. 
For simplicity we choose the branch cuts of all MFs to be in the same direction and number the MFs 
based on their position orthogonal to this direction, such that when exchanging MFs $i$ and $i + 1$ in a clockwise manner, 
vortex $i$ crosses \emph{only} the branch cut of vortex $i+1$, and no other vortices cross any branch cuts (crossing the same 
branch cut twice in different directions is equivalent to not crossing any branch cuts at all). 
Note that MFs which are not neighbors can always be exchanged through a sequence of neighbor exchanges.
Consider now the effect of braid operations on the number states 
\begin{eqnarray}
\label{eq:exchange12}
	B_{12} |00 \rangle &=& \frac{1}{\sqrt{2}} \left( 1 + i \right) |00\rangle, \\
\label{eq:exchange23}
	B_{23} |00 \rangle &=& \frac{1}{\sqrt{2}} \left( |00\rangle + i |11\rangle \right), \\
\label{eq:exchange34}
	B_{34} |00 \rangle &=& \frac{1}{\sqrt{2}} \left( 1 + i \right) |00\rangle,
\end{eqnarray}
with analogous results for the other number states. Note especially that $B_{23}$, which involves MFs from different fermions, produces
a superposition state of different number states. However, the \emph{total} parity ($n_1 + n_2$ being even or odd) of each state in the 
superposition must be the same.

With four MFs we can also demonstrate the non-abelian nature of braid operations (with two MFs there is only one possible exchange operation).
In general, two braid operations commute whenever they do not involve any of the same MFs, $[B_{12}, B_{34}] = 0$. This is easy to believe 
on physical grounds, as there is no reason that the exchange of MFs three and four should care about whether MFs one and two have been 
exchanged. However, whenever two exchanges involve some of the same MFs, the braid operators do not commute
\begin{eqnarray}\label{eq:noncommuting}
	\left[ B_{i-1,i}, B_{i, i+1} \right] &=& \gamma_{i-1} \gamma_{i+1}.
\end{eqnarray}
Equation~(\ref{eq:noncommuting}) expresses the non-abelian exchange statistics of MFs.

\begin{figure}[h!]
  \includegraphics[height=0.45\linewidth]{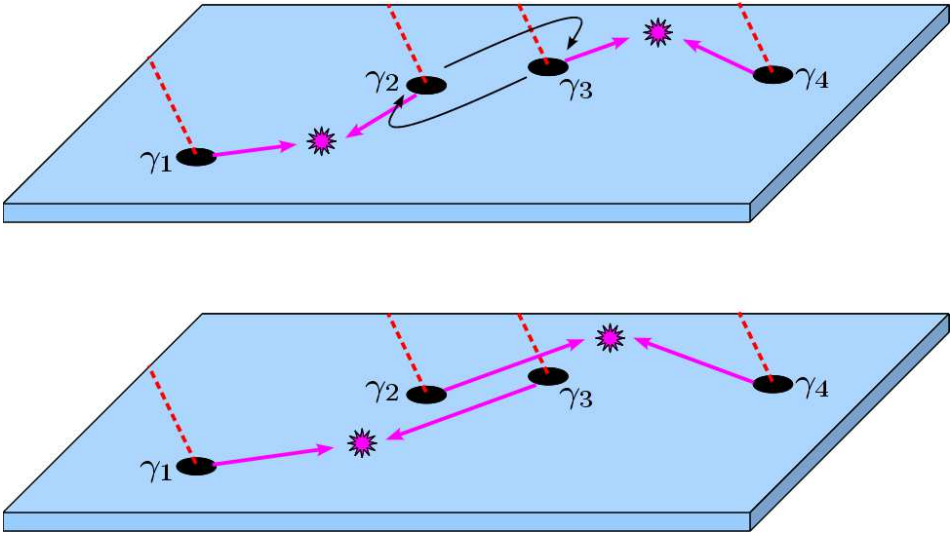}
	\caption{\label{fig:exchange2}
	Sketch demonstrating two equivalent sets of operations. In the upper panel, MFs 2 and 3 are first exchanged (black arrows), 
	then the nearest neighbor MFs are brought together (magenta arrows) and the states of the corresponding fermions 
	are measured.  
	In the lower panel, there is no exchange, but instead we directly measure the fermions formed by pairing next-nearest 
	neighboring MFs ($1 + 3$ and $2 + 4$).}
\end{figure}

At this point the attentive reader might be slightly upset by a simple fact we have neglected. Namely the question of what 
exactly qualifies as an exchange operation. If we define an exchange operation as bringing one vortex exactly to the old 
position of another vortex, and vice versa, there is no problem. But clearly this is not possible in reality and certainly 
goes against the idea of robust topological quantum information processing to be discussed below. (In networks of 1D wires it 
is somewhat easier to find a satisfying definition of particle exchanges~\cite{Alicea10b}.) 
Mathematically, the exchange process 
happens when the branch cut is crossed, but since this is arbitrarily defined, it is not a good definition either. Physically, 
the solution to this problem is that there is no measurable effect of the exchange process, unless it is followed by one of the 
two MFs involved in the exchange being joined with a third MF to perform a measurement of the state of the fermion formed by 
this pair. 
This is demonstrated in Fig.~\ref{fig:exchange2}. 
In the upper panel, two neighboring MFs are first exchanged, 
which is followed by measuring the fermionic states formed by pairing the nearest neighbor MFs.  
In the lower panel, on the other hand, we do not exchange the MFs, but instead measure directly the fermionic states 
formed by pairing next-nearest neighboring MFs.
Both these operations give the same result for the measurements of the fermionic states and are therefore equivalent.  
Therefore, a "computation" can be defined either as a set of exchanges, or by defining the combinations of pairs that are being 
measured at the end, which removes the ambiguity in defining an exchange process.

Another way of seeing this is to note that the result of a set of exchange operations depends on the topology of the "world lines" 
of the MFs (or other particles). The world lines start when pairs of MFs are created from the vacuum, cross each other when MFs are exchanged, and 
end when pairs of MFs are measured (fused). Only sets of world lines which are topologically distinct can lead to different results, which defines the 
so-called braid group, of which the MF exchange statistics is one possible representation (fermions, bosons, and abelian anyons in general are the 
more trivial one-dimensional representations). Non-abelian representations of the braid group exist only in 2D (networks of 1D wires are also 
effectively 2D). Ref.~\cite{Nayak08rev} provides a more complete mathematical discussion of these issues.

\subsection{Majorana qubits and topological quantum computation}
We saw above that parity conservation prevented braiding operations from changing the state of a system with only two MFs. 
For this reason, the two-level system spanned by the number operator $n_1$ is not suitable to use as a qubit. 
To define a (topological) Majorana qubit, we should therefore use four MFs, meaning two normal fermions~\cite{Bravyi06}, and consider 
the case of fixed parity (even or odd) of the total number of fermions. 
Let us consider the even parity subspace and define a 
qubit by $|\bar{0} \rangle \equiv |00\rangle$, $|\bar{1} \rangle \equiv |11\rangle$. 

In the basis $\{\bar{0},\bar{1}\}$, the Pauli matrices can be represented in terms of products of Majorana operators
\begin{eqnarray}
\label{eq:pauliMFz}
	-i \gamma_1 \gamma_2 = \sigma_z, -i \gamma_3 \gamma_4 = \sigma_z \\
\label{eq:pauliMFx}
	-i \gamma_2 \gamma_3 = \sigma_x, \\
\label{eq:pauliMFy}
-i \gamma_1 \gamma_3 = \sigma_y, -i \gamma_2 \gamma_4 = \sigma_y
\end{eqnarray}
which is seen by calculating the corresponding matrix elements. 
In the standard representation with $|\bar{0}\rangle$ and $|\bar{1}\rangle$ being respectively the north and south poles of the block 
sphere, we can then identify the different braids with single-qubit rotations
\begin{eqnarray}
\label{eq:rotz}
	B_{12} = B_{34} = e^{-\frac{i\pi}{4}\sigma_z}, \\
\label{eq:rotx}
	B_{23} = e^{-\frac{i\pi}{4} \sigma_x},
\end{eqnarray}
Thus, by braiding operations we can only perform single-qubit rotations by an angle $\pi/2$. 

When considering a multi-qubit setup, the most obvious choice is to define each qubit in terms of four MFs~\cite{Bravyi06}. 
However, it is not possible to construct a two-qubit gate based on braiding operations which is able to create entanglement 
between two such qubits. In addition, this choice does not use all the degrees of freedom offered by the system.
Even with conservation of the parity of the total number of fermions, a system with $2N$ MFs has in principle enough degrees of freedom to 
store $N-1$ qubits. Braiding-based gates acting on such "overlapping" qubit systems have been considered in Ref.~\cite{Georgiev06}.
No matter how the Majorana-based qubits are defined, braiding operations can only explore a tiny fraction of the total Hilbert space 
and are insufficient for universal quantum computation, which can, however, be 
achieved by including also non-protected operations~\cite{Kitaev03, Nayak08rev}, or by coupling Majorana qubits to other
qubit systems~\cite{Hassler10, Sau10b, Jiang11, Bonderson11, Leijnse11top}.

The advantage of Majorana-based qubits is that they keep the quantum information encoded in delocalized fermionic states. 
Therefore, they are expected to be robust against most sources of decoherence which do not couple simultaneously to more 
than one Majorana mode, i.e., decoherence requires perturbations of the form $\gamma_i \gamma_j$, which are suppressed 
when MFs $i$ and $j$ are spatially separated. 
An exception is processes which change the total parity of the superconductor, e.g., by electrons tunneling into a 
Majorana mode~\cite{Leijnse11, Budich12, Rainis12}. 
Such perturbations involve a single Majorana operator (they are $\propto\gamma_i$)
and are not suppressed by 
keeping the MFs spatially separated. In fact, this phenomena, known as quasiparticle 
poisoning~\cite{Mannik04, Aumentado04}, is a well-known problem in conventional (non-topological) superconducting qubits.

\section{Proximity-induced superconductivity in spin-orbit semiconductors}
Having seen how Majorana fermions appear in spinless $p$-wave (or $p_x \pm ip_y$-wave) superconductors, 
we will now investigate how such exotic pairing can be engineered using more readily available ingredients.
We start, however, by considering the generic effects of proximity-induced superconductivity.

The system we have in mind is a $D$-dimensional semiconductor, where $D=1$ (wire) or $D=2$ (quantum well). 
Neglecting electron--electron interactions (or including them in a mean-field manner), the system is described 
by the Hamiltonian
\begin{eqnarray}\label{eq:H0}
	\mathcal{H}_0 &=& \sum_{\sigma = \uparrow,\downarrow}\int d^D r \; \Psi_\sigma^\dagger(\mathbf{r}) H_0(\mathbf{r}) \Psi_\sigma(\mathbf{r}),
\end{eqnarray}
where the first quantization single-particle Hamiltonian is given by
\begin{eqnarray}\label{eq:H0r}
	H_0(\mathbf{r}) &=& \frac{\mathbf{p}^2}{2 m} - \mu + V(\mathbf{r}) + 
			    \alpha \left( \mathbf{E}(\mathbf{r}) \times \mathbf{p} \right) \cdot \bar{\sigma} + 
			    \frac{1}{2} g \mu_B \mathbf{B}(\mathbf{r}) \cdot \bar{\sigma},
\end{eqnarray}
where $m$ is the effective electron mass, $\mathbf{B}$ is an applied magnetic field, $\mu_B$ is the Bohr magneton, $g$ is the Land\'{e} $g$ factor, 
and $\bar{\sigma}$ is a vector of Pauli matrices.  
The spin-orbit interaction with strength $\alpha$ has been written in the most general form in terms of the electric field 
$\mathbf{E}$ felt by the valence electrons, and can involve both Rashba and Dresselhaus terms.   

If a good interface is made between a semiconductor and a superconductor, electrons can tunnel between these two systems. 
The effect is that the electrons in the semiconductor feels an effective "proximity-induced" superconducting pairing 
field~\cite{deGennes64, Doh05, vanDam06}. The strength of this pairing field depends on the details of the semiconductor and 
superconductor, as well as the interface. We do not attempt to make an accurate microscopic model, but instead include 
the pairing effect in the semiconductor by the phenomenological Hamiltonian
\begin{eqnarray}\label{eq:HS}
	\mathcal{H}_S &=& \int d^D r \; d^D r' \; \Psi_\downarrow(\mathbf{r}) \Delta(\mathbf{r}, \mathbf{r'}) 
		\Psi_\uparrow(\mathbf{r'}) + h.c.,
\end{eqnarray}
where $\Delta(\mathbf{r}, \mathbf{r'})$ is the pairing potential. 
The pairing symmetry is inherited from the superconductor, which we have here assumed to be $s$-wave, 
inducing singlet pairing between spin-up and spin-down electrons. 

When dealing with superconducting systems, it is standard practice to include both the electrons and the holes explicitly 
by introducing so-called Nambu spinors
\begin{eqnarray}\label{eq:nambu}
	\bar{\Psi}(\mathbf{r}) = \left( \begin{array}{c}\Psi_\uparrow(\mathbf{r}) \\
				  			\Psi_\downarrow(\mathbf{r}) \\
							\Psi_\downarrow^\dagger(\mathbf{r}) \\
							-\Psi_\uparrow^\dagger(\mathbf{r}) \end{array} \right).
\end{eqnarray}
Matrices acting on the Nambu spinors must have dimension $4\times4$ and we introduce Pauli matrices $\tau_i$, 
similar to $\sigma_i$, but acting in electron-hole space. Matrices such as $ \tau_j \otimes \sigma_i$ then have the 
appropriate $4\times4$ structure.
The total Hamiltonian can be written as
\begin{eqnarray}\label{eq:H}
	\mathcal{H} &=& \mathcal{H}_0 + \mathcal{H}_S \nonumber \\ 
		    &=& \frac{1}{2} \int d^D r \; d^D r' \; 
			\bar{\Psi}^\dagger(\mathbf{r}) \left[ \bar{H}_0(\mathbf{r}) \delta(\mathbf{r} - \mathbf{r'}) 
			 +  \bar{\Delta}(\mathbf{r}, \mathbf{r'}) \right] \bar{\Psi}(\mathbf{r}),  
\end{eqnarray}
where
\begin{eqnarray}
\label{eq:H0bar}
	\bar{H}_0(\mathbf{r})  &=&  \left( \begin{array}{cc} H_0(\mathbf{r}) & \hat{0}_\sigma \\ \hat{0}_\sigma & - \sigma_y H_0^* (\mathbf{r}) \sigma_y \\ 
				  \end{array} \right), \\ \nonumber \\
\label{eq:Deltabar}
        \bar{\Delta}(\mathbf{r}, \mathbf{r'})  &=&  \left( \begin{array}{cc}  
				  \hat{0}_\sigma & \Delta^*(\mathbf{r}, \mathbf{r'}) \hat{1}_\sigma \\ 
				  \Delta(\mathbf{r}, \mathbf{r'}) \hat{1}_\sigma & \hat{0}_\sigma \\ \end{array} \right).
\end{eqnarray}
Note that these are $4\times4$ matrices since $H_0(\mathbf{r})$ is a $2\times2$ matrix and $\hat{0}_\sigma$ and $\hat{1}_\sigma$
denotes respectively the zero and unit matrices in spin space.
The term $-\sigma_y H_0^* (\mathbf{r}) \sigma_y$ in Eq.~(\ref{eq:H0bar}) denotes the time-reversal of $H_0(\mathbf{r})$ and appears since 
holes are time-reversed electrons. (Note, however, that we have not introduced any new physics with the matrices in 
Eqs.~(\ref{eq:H0bar})--(\ref{eq:Deltabar}), they are simply defined to give 
the correct total Hamiltonian, given by Eq.~(\ref{eq:H0}) + Eq.~(\ref{eq:HS}).)

The quasiparticle excitations of a superconducting system are given by solving the Bogoliubov-de Gennes equations for the eigenstates 
$\bar{\psi}_i(\mathbf{r})$ (which are also four-component spinors)
\begin{eqnarray}\label{eq:BdG}
	\bar{H}_0 (\mathbf{r}) \bar{\psi}_i(\mathbf{r}) + \int d^D r' \; \bar{\Delta}(\mathbf{r}, \mathbf{r'}) \bar{\psi}_i(\mathbf{r'}) = 
		E_i \bar{\psi}_i(\mathbf{r}),
\end{eqnarray}
and the diagonalized Hamiltonian becomes
\begin{eqnarray}
\label{eq:Hdiag}
	\mathcal{H} &=& \frac{1}{2} \sum_i  E_i \; \Psi_i^\dagger \Psi_i, \\
\label{eq:eigenop}
	\Psi_i &=& \int d^D r \; \bar{\psi}_i (\mathbf{r}) \cdot \bar{\Psi}(\mathbf{r}).	
\end{eqnarray}

As was mentioned above, the Nambu spinors were only introduced to provide a convenient matrix representation of the Hamiltonian.
However, by explicitly including the hole states and thus doubling the dimension of the Hamiltonian, we have also artificially 
doubled the number of eigenstates. Therefore, there must be some symmetry relation between the eigenstates such that the number 
of independent solutions remains the same. This symmetry is electron-hole symmetry, which is expressed through the operator
\begin{eqnarray}\label{eq:Pop}
	P =  \tau_y \otimes \sigma_y K = \left( \begin{array}{cccc} 0 & 0 & 0 & -1 \\ 0 & 0 & 1 & 0 
					                         \\ 0 & 1 & 0 & 0 \\ -1 & 0 & 0 & 0\end{array} \right) K,
\end{eqnarray}
where $K$ is the operator for complex conjugation. It can then be verified that  
\begin{eqnarray}\label{eq:ehsymm}
	P \bar{H}_0 (\mathbf{r}) P^\dagger &=& -\bar{H}_0 (\mathbf{r}), \\
	P \bar{\Delta} (\mathbf{r}, \mathbf{r'}) P^\dagger &=& -\bar{\Delta} (\mathbf{r}, \mathbf{r'}).	 
\end{eqnarray}
This means that if $\bar{\psi}_{i}(\mathbf{r})$ and $\Psi_i$ are solutions with positive energy $E_i$, then there 
are also solutions $\bar{\psi}_{j}(\mathbf{r})$ and $\Psi_j$ with energy $E_j = -E_i$ which satisfy 
\begin{eqnarray}
\label{eq:ehwavefuncsymm}
	\bar{\psi}_{j}(\mathbf{r}) = P \bar{\psi}_{i}(\mathbf{r}), \\
\label{eq:ehopsymm}
	\Psi_{j} = \Psi_{i}^\dagger. 
\end{eqnarray}
In other words, creating a quasiparticle with energy $E$ or removing one with energy $-E$ are identical operations.

It is possible to engineer the parameters in the Hamiltonian~(\ref{eq:H}) such that it resembles a spinless $p$-wave 
or $p_x \pm ip_y$ superconductor and has eigenstates which are MFs. We will discuss how to accomplish this in the following section, 
but can already now derive a few of the generic properties Majorana solutions to Eq.~(\ref{eq:BdG}) must have, 
\emph{if} they exist.
Let us assume that there is a solution to Eq.~(\ref{eq:BdG}) which corresponds to a Majorana operator 
$\gamma_i \equiv \Psi_i = \Psi_i^\dagger$. 
According to Eq.~(\ref{eq:ehopsymm}), this is only possible at $E=0$, so a MF 
is a zero-energy solution to the Bogoliubov-de Gennes equations. 
The corresponding real-space Majorana spinor
satisfies $\bar{\psi}_M(\mathbf{r}) = P \bar{\psi}_M(\mathbf{r})$, and its most general form is therefore  
\begin{eqnarray}
\label{eq:realspaceMF}
	\bar{\psi}_M(\mathbf{r}) = \left( \begin{array}{c} f(\mathbf{r}) \\ g(\mathbf{r}) \\ g^*(\mathbf{r}) \\ -f^*(\mathbf{r}))
			     \end{array}\right).
\end{eqnarray}

Let us now assume that the Hamiltonian respects time-reversal symmetry, $H = T H T^\dagger$, where
the time-reversal operator is $T = i \hat{1}_\tau \otimes \sigma_y K$. 
Then, if $\bar{\psi}_i(\mathbf{r})$ is an eigenstate, also $T \bar{\psi}_i(\mathbf{r})$ must be 
an eigenstate with the same energy. This is nothing else than Kramer's theorem for a superconductor.
Specifically for MFs, the Kramer's partner to the Majorana spinor $\bar{\psi}_M(\mathbf{r})$ in 
Eq.~(\ref{eq:realspaceMF}) is
\begin{eqnarray}
\label{eq:realspaceMFpartner}
	\bar{\psi}_{M'}(\mathbf{r}) = \left( \begin{array}{c} g^*(\mathbf{r}) \\ -f^*(\mathbf{r}) \\ -f(\mathbf{r}) \\ -g(\mathbf{r})
			     \end{array}\right).
\end{eqnarray}
Importantly, the probability densities associated with each Kramer's pair are identical
\begin{eqnarray}\label{eq:kramersspatial}
	|\bar{\psi}_M(\mathbf{r})|^2 = |\bar{\psi}_{M'}(\mathbf{r})|^2 = 2|g(\mathbf{r})|^2 + 2|f(\mathbf{r})|^2.
\end{eqnarray}
Thus, if we want to create spatially isolated MFs (not accompanied by a Kramer's partner with identical probability density), 
we need a Hamiltonian which breaks time-reversal symmetry.

We note that it is possible for pairs of MFs to remain at zero energy in the presence of time-reversal symmetry even if they are 
not spatially separated. For a full classification of different topological classes, see e.g.,~\cite{Ryu10, Fulga11}. 

\section{Induced $p$-wave-like gap in semiconductors}
The strategy is now to engineer the parameters in the semiconductor Hamiltonians~(\ref{eq:H0})--(\ref{eq:HS}), such that
the total Hamiltonian~(\ref{eq:H}) is close enough to that of a spinless $p$-wave 
superconductor (Eq.~(\ref{eq:pgap1d})) 
to also host MFs (we focus here on the simplest case of a 1D nanowire, but the case of a 2D quantum well is rather similar). 
The precise meaning of "close enough" is here that the two Hamiltonians can be continuously transformed into each other, without 
ever closing the gap. In this case Eq.~(\ref{eq:H}) and Eq.~(\ref{eq:pgap1d}) describe topologically 
equivalent systems. Since the presence of spatially separated MFs
is a topological property of the system~\cite{Hasan10}, they should then appear also as solutions to Eq.~(\ref{eq:H}).

\begin{figure}[h!]
  \includegraphics[height=0.42\linewidth]{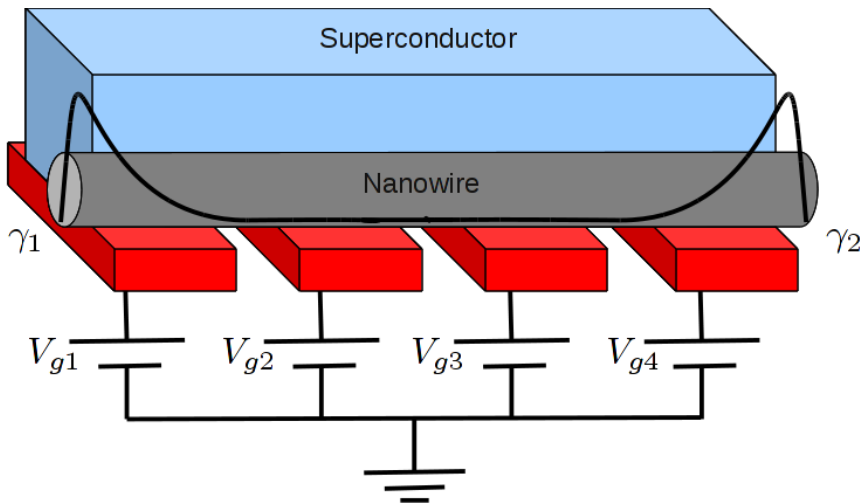}
\begin{centering}	
	\caption{\label{fig:wire}
	Sketch of setup for engineering topological superconductivity in a 1D nanowire. 
	The nanowire (e.g., InAs or InSb) with strong spin-orbit coupling is proximity coupled to 
	a bulk $s$-wave superconductor (e.g., Nb or Al). A set of gate electrodes are used to control the chemical potential inside the wire
	and bring it into the topological regime. MFs then form at the ends of the wire. The weight of the Majorana wavefunction decays 
	exponentially inside the wire, indicated here in black (this is just an approximate form, the real wavefunction depends on the details 
	and often exhibit oscillations). }
\end{centering}
\end{figure}

A 1D semiconducting wire with strong spin-orbit coupling has been put forward as an experimentally attractive setting in which 
to induce topological superconductivity. 
We follow here closely the proposals originally put forward in Refs.~\cite{Oreg10, Lutchyn10}. The experimental geometry is 
sketched in Fig.~\ref{fig:wire}.  
The nanowire is proximity-coupled to a $s$-wave superconductor and exposed to an external magnetic field (not shown). The chemical potential 
of the wire is controlled by a set of gate electrodes. The wire is assumed to be long enough that we can ignore size quantization along the 
wire direction and thin enough that the 1D subbands are well separated on the relevant energy scales. For simplicity, we also assume that 
the chemical potential can be tuned to a regime where only a single 1D subband is occupied (MFs can also be found in multi-channel 
wires~\cite{Potter10, Stanescu11, Fulga11} provided that the channel number is odd and the effective width of the wire is smaller 
than the superconducting coherence length). The Hamiltonian is a special case of Eq.~(\ref{eq:H0r})
\begin{eqnarray}\label{eq:Hwire}
	H_{0}(x) = \frac{k_x^2}{2 m} - \mu + \tilde{\alpha} k_x \sigma_y + \frac{1}{2} \tilde{B} \sigma_z,
\end{eqnarray}
where we have taken $\hbar = 1$. $\tilde{\alpha} = \alpha E_\perp$, with $E_\perp$ the electric field perpendicular to the wire 
direction, is the strength of the spin-orbit field originating 
from Rashba spin-orbit coupling and $\tilde{B} = g \mu_B B$ is the Zeeman field. Note that the direction of the electric field 
in a realistic setup is unknown due to the presence of the 
superconductor and electric gates. However, since the spin-orbit field is given by a cross product of the electric field and the momentum, 
it is orthogonal to the wire and we choose it to 
be along the $y$-direction. Only the magnetic field component orthogonal to the spin-orbit field will help to induce topological 
superconductivity and we choose the field to be along the $z$-axis.
We assume the proximity-induced pairing field to be homogeneous and to couple only electrons at the same position, 
$\Delta(x,x') = \Delta \delta(x - x')$ in Eq.~(\ref{eq:HS}).

\begin{figure}[h!]
  \includegraphics[height=0.65\linewidth]{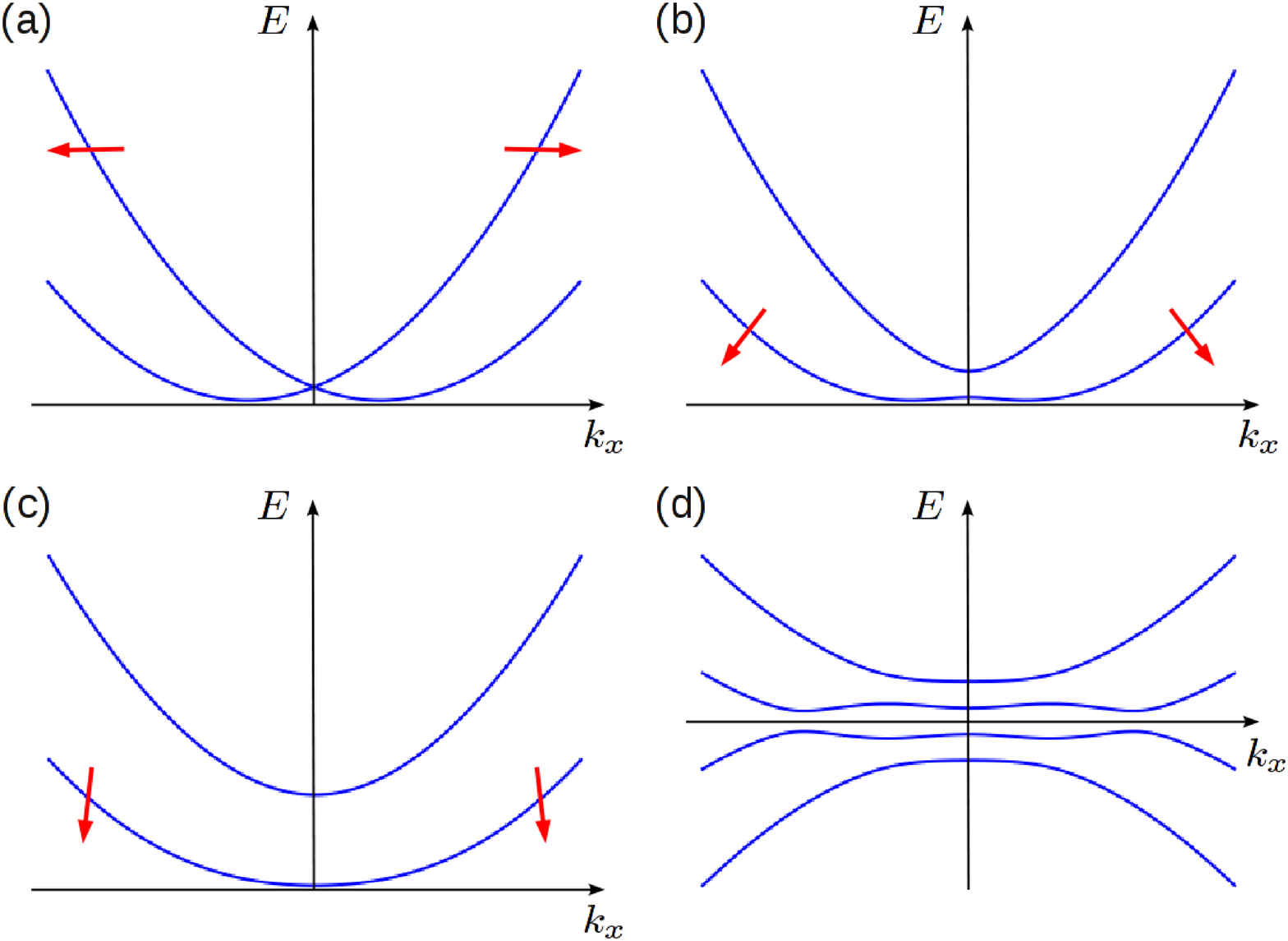}
\begin{centering}	
	\caption{\label{fig:bands}
	Bandstructure, $E(k_x)$, of a spin-orbit coupled nanowire with applied magnetic field, $\tilde{B}$. 
	(a) $\tilde{B}=0$. The two parabolic bands are split depending on their spin polarization (red arrows) 
	along the axis of the spin-orbit field. 
	(b) Small $\tilde{B} > 0$. The field opens a gap at $k_x = 0$ and thereby a region without spin degeneracy. 
	Each band holds only a single spin projection, but the direction of the spin polarization depends on the momentum. 
	(c) Large $\tilde{B}$. The effectively spinless regime grows with magnetic field and also forces the spins within each band to 
	increasingly align either parallel or anti-parallel to the field. 
	(d) Same as (c), but with proximity-induced superconductivity, $\Delta > 0$. Here we plot all the solutions of the Bogoliubov-de 
	Gennes equations (Eq.~\ref{eq:BdG}), giving twice the number of solutions. Note the symmetry between positive and negative 
	$E$ due to the particle-hole symmetry.  
	}
\end{centering}
\end{figure}

The eigenspectrum of Eq.~(\ref{eq:Hwire}) is shown in Fig.~\ref{fig:bands} as a function of the momentum along the wire. 
When $\tilde{B}=0$, see Fig.~\ref{fig:bands}(a), the spin-orbit coupling shifts the two parabolic bands depending on their spin polarization
along the axis of the spin-orbit field. However, at any given energy there is still spin degeneracy since time-reversal symmetry is
not broken.
Switching on a small $\tilde{B}$, see Fig.~\ref{fig:bands}(b), the crossing at zero momentum turns into an anti-crossing. 
The lower band has a double-well shape and the gap 
to the upper band is determined by $\tilde{B}$ (only the component of the magnetic field orthogonal to the spin-orbit field opens up a gap).
The Hamiltonian~(\ref{eq:Hwire}) is easily diagonalized, resulting in
\begin{eqnarray}\label{eq:bands}
	E_{\pm}(k_x) = \frac{k_x^2}{2m} - \mu \pm \sqrt{(\tilde{\alpha} k_x)^2 + \tilde{B}^2},
\end{eqnarray}
where $E_{+}(k_x)$ and $E_{-}(k_x)$ correspond to the upper and lower bands, respectively.
Inside the gap, there is only one effective spin direction (although this direction depends on 
momentum). Therefore, if $\mu$ is placed inside the gap, spinless superconductivity can be induced by the proximity effect. Note that without the 
$k_x$-dependence of the spin direction, it would not be possible to induce superconductivity by proximity with a $s$-wave superconductor, 
since the pairing term in Eq.~(\ref{eq:HS}) only couples the components of the spins which are anti-parallel. 
A larger $\tilde{B}$, as in Fig.~\ref{fig:bands}(c), increases the gap between the bands. This makes the effectively spinless regime larger and
therefore provides a larger window in which to place the chemical potential. 
This is important if disorder causes the the chemical potential to vary over the length of the wire. 
However, the larger field also enforces an increased
alignment of the spins within each band and therefore makes it harder to induce superconductivity.

Next we switch on also the proximity-induced superconducting pairing, $\Delta > 0$, see Fig.~\ref{fig:bands}(d). 
Here we plot all the solutions of the Bogoliubov-de Gennes equations, Eq.~(\ref{eq:BdG}), which doubles the number of bands
(there are still only two independent solutions since particle-hole symmetry forces the positive and negative energy solutions to be 
identical).
For small $\Delta$, the superconducting state is topological and associated with Majorana edge states, provided that 
the chemical potential is placed within the spinless regime, $|\tilde{B}| > |\mu|$. 
The gap at zero momentum decreases with increasing $\Delta$ and closes completely when $|\tilde{B}| = \sqrt{\Delta^2 + \mu^2}$. 
For larger values of $\Delta$ the gap opens again, but now in a non-topological superconducting state. 
The criteria for topological superconductivity is therefore
\begin{eqnarray}\label{eq:topcrit}
	|\tilde{B}| > \sqrt{\Delta^2 + \mu^2}.
\end{eqnarray}
The phase transition between the topological and non-topological superconducting states can \emph{only} take place at the
point where the gap closes. Therefore, to show that the criteria~(\ref{eq:topcrit}) indeed guarantees topological superconductivity, 
it is sufficient to consider a certain limit of the topological phase and there provide a mapping to the spinless $p$-wave 
superconducting wire in Eq.~(\ref{eq:pgap1d}) (and thereby to Kitaev's chain described by Eq.~(\ref{eq:kitaevH})). 

Following Ref.~\cite{Alicea10b}, we consider the limit $|\tilde{B}| \gg E_{so}, |\Delta|$ and $\mu = 0$, where 
$E_{so} = m \tilde{\alpha}^2 / 2$ is the spin-orbit energy.
With such a large magnetic field, the spins within each bands are nearly completely polarized and since the gap is large 
we can consider a single-band model and ignore the higher band. Let $\Psi_{-}(x)$ be the annihilation operator 
for electrons at position $x$ in the lower band of the wire. The bands are nearly spin polarized in the direction 
of the magnetic field and therefore $\Psi_-(x) \approx \Psi_\downarrow (x)$. However, we need to take into account leading order 
corrections due to spin-orbit couplings, since the induced superconductivity is $\propto \Psi_\downarrow \Psi_\uparrow (x)$. 
The effective Hamiltonian for the lower band with induced superconcductivity then becomes
\begin{eqnarray}\label{eq:Heff}
	\mathcal{H}_\mathrm{eff} = \int dx \; \left[ \Psi^\dagger_-(x)\left( \frac{k_x^2}{2 m} - \frac{1}{2} \tilde{B} \right)\Psi_-(x) 
				    + i \frac{\tilde{\alpha}}{\tilde{B}} \Delta \Psi_-(x) k_x \Psi_-(x) \right].
\end{eqnarray}
This is equivalent to the Hamiltonian~(\ref{eq:pgap1d}). 
Note that the effective strength of the superconducting pairing term has been reduced by a factor $\tilde{\alpha} / \tilde{B}$, 
corresponding to the polarization of the spin within the lower band, suppressing superconducting pairing. 

From the above discussion and from Eq.~(\ref{eq:topcrit}), it may seem like an arbitrarily small proximity-induced gap is 
sufficient for topological superconductivity. However, the stability of the superconducting phase depends also on the size 
of the gap at finite momentum (see Fig.~\ref{fig:bands}(d)), which is $\propto \Delta$, see Eq.~(\ref{eq:Heff}). 
If the gap is too small, the topological 
superconducting phase, although reached in the idealized model considered here, will in reality be destroyed e.g., by finite 
temperature or disorder. 
Clearly it is important to induce a significant Zeemann splitting in the wire. Since this has to be done without destroying 
superconductivity, a semiconductor with a large $g$-factor is desirable. A large Zeeman splitting ensures a large gap at zero 
momentum, which is important since the chemical potential must be placed within the gap. 
Finally, it is important to have a large spin orbit coupling, since the gap at finite momentum is otherwise suppressed by the 
magnetic field, see Eq.~(\ref{eq:Heff}). Furthermore, the sensitivity to disorder increases when time-reversal 
symmetry is strongly violated for electrons close to the Fermi level~\cite{Potter11, Sau11c}, meaning when the spins at $\pm k_x^F$ are 
almost aligned. To avoid this, and keep robostness against disorder, one should not allow the ratio $\tilde{B} / E_{so}$, 
to become too large.
Suitable candidate materials for nanowires, with large $g$-factors and strong spin-orbit coupling, are for example InAs and InSb.

\section{Conclusions and outlook}
In this short review article we have tried to give a pedagogical introduction to the exciting field of topological 
superconductivity and MFs in condensed matter systems. 
We saw that MFs, being equal superpositions of electrons and holes, can appear as Bogoliubov quasiparticles in effectively 
spinless superconductors. Their peculiar non-abelian (non-commutative) exchange statistics was discussed, and the basic concepts 
of Majorana qubits and topological quantum computation was introduced. Finally, we discussed the possibility to realize topological 
superconductivity by bringing semiconductors with strong spin-orbit coupling into proximity with standard $s$-wave superconductors 
and exposing them to a magnetic field. 

While this manuscript was being finalized, a report of the possible observation of MFs in 1D nanowires was published~\cite{Mourik12}, 
and around the same time several other groups reported observations which can be interpreted as signatures of 
Majoranas~\cite{Williams12, Rokhinson12, Deng12} (for space reasons we have here not discussed the issue of Majorana detection, 
see e.g.,~\cite{Beenakker11rev, Alicea12rev}).
Clearly the types of systems investigated in the hunt for Majoranas contain a lot of new and exciting physics, but which experimental
observation are indeed genuine Majorana sightings remains to be seen. We also mention that in a real system, with a finite size 
and interactions which may not be well described within a mean-field picture, there is not necessarily a perfectly clear and 
unique definition of a MF. 

In any case, once MFs can be reproducibly realized and detected in the lab, the work has only started. 
Further studies will have to investigate their properties in detail, and fabrication and measurement techniques will 
have to be perfected. No doubt, there will also be a need for additional theoretical work to understand the experimental findings.
On a longer timescale, the goal is of course to be able to control and manipulate quantum information stored in 
Majorana-based qubit systems. If there will be a useful technological application at the end of this long road is too
early to predict, but there is certainly a lot of interesting and new physics to explore.

\section*{References}

\bibliographystyle{unsrt}

\end{document}